\documentclass[acmsmall,nonacm,screen]{acmart}
\acmJournal{PACMPL}
\acmNumber{CONF} 
\acmYear{2023}
\acmMonth{1}
\acmDOI{} 
\startPage{1}

\setcopyright{none}

\usepackage{header}

\begin{document}

\title{A Two-Level Linear Dependent Type Theory}

\author{Qiancheng Fu}
\affiliation{
  \institution{Boston University}
  \city{Boston}
  \state{MA}
  \country{USA}
}
\email{qcfu@bu.edu}

\author{Hongwei Xi}
\affiliation{
  \institution{Boston University}
  \city{Boston}
  \state{MA}
  \country{USA}
}
\email{hwxi@bu.edu}

\begin{abstract}
  We present a type theory combining both linearity and dependency by stratifying typing rules into a level for logics and a level for programs. The distinction between logics and programs decouples their semantics, allowing the type system to assume tight resource bounds. A natural notion of irrelevancy is established where all proofs and types occurring inside programs are fully erasable without compromising their operational behavior. Through a heap-based operational semantics, we show that extracted programs always make computational progress and run memory clean. Additionally, programs can be freely reflected into the logical level for conducting deep proofs in the style of standard dependent type theories. This enables one to write resource safe programs and verify their correctness using a unified language. 
\end{abstract}


\maketitle

\section{Introduction}
Constructing programs and proving their properties are primary functions of computer science. It did not take long for researchers to realize that a deep connection exists between these activities. Proofs can be viewed as programs mapping assumptions to conclusions and programs can be viewed as proofs of the propositions induced by types. This realization lead to the Curry-Howard correspondence viewpoint which identifies proofs and programs as the same mathematical objects. Dependent type theories~\cite{martinlof,cc,ecc} make this connection even more concrete by expressing proofs and programs using a single unified language.

Despite these theoretical observations, the generality of the Curry-Howard correspondence is still a contentious subject. It does not take much time to notice that the focal points of theorem proving and program construction are different. On the one hand, the value of a proof lies in its ability to assert the validity of a proposition by being assigned the type corresponding to this proposition. In other words, the reduction behavior (computation) of a proof is not as important as the fact that it is well-typed. Proofs utilizing classical axioms such as the law of excluded middle or the axiom of choice may not reduce at all. On the other hand, the value of a program lies in its ability to perform computation and transform the world around it (sometimes irreversibly). Many programming languages such as C do not have clear analogs in the proof theoretic world and yet have greatly impacted modern society solely through their computational capabilities.

As a step towards bridging the gap between proofs and programs, we introduce the dependent type theory TLL whose typing rules are stratified into a logical level and a program level. Now that proofs and programs are no longer bound together in a monolithic type system, typing rules can be refined to characterize their subjects of interest more precisely. At the logical level, the typing rules here are concerned with the formation of propositions (types) and their proofs (terms). Overall, the logical level is similar to Martin-L\"{o}f type theory~\cite{martinlof} and enjoys many of the same meta theoretical properties. For the typing rules at the program level, they support an intrinsic notion of resources inspired by Linear Logic~\cite{girard} to encode irreversible real world interactions. Due to the fact that the operational semantics of the two levels can be decoupled, we choose a call-by-value style semantics for programs as this allows the program typing rules to assume tight resource bounds. Additionally, TLL types are allowed to depend on linearly typed programs in a computationally irrelevant manner. This enables one to reason about resources hypothetically without actually consuming them.

In order to show that TLL guarantees productive programs and safe resource usage, we develop a type directed erasure procedure and a heap semantics inspired by Turner and Wadler~\cite{turner99}. During erasure, the syntax tree of a well-typed program is stripped of all type annotations and computationally irrelevant terms. The program extracted from erasure is then evaluated using our heap semantics. As the program evaluates, heap cells are dynamically allocated and freed when linearly typed values are constructed and consumed. We prove that our calculus is sound with regards to this erasure procedure and heap semantics, ensuring evaluation progress and safe memory usage at runtime.

All lemmas and theorems reported in this paper are formalized and proven correct in Coq~\cite{coq}. We also implement a compiler in OCaml that compiles TLL programs into C. Proofs, source code and example programs are available in our git repository~\footnote{\url{https://github.com/qcfu-bu/TLL-arxiv-repo}}.

In summary, we make the following contributions:

\begin{itemize}
  \item First, we design TLL, a two-level linear dependent type system. By stratifying the typing rules into a logical level and a program level, we are able to characterize proofs and programs more precisely.
  \item Second, we study the meta-theoretical properties of the two levels. We show that the logical level exhibits qualities such as confluence and strong normalization that make it suitable for logical reasoning.
  \item Furthermore, we design an erasure procedure and heap semantics that model the behavior of programs at runtime. Using this semantics, we show programs extracted from erasure run memory clean.
  \item The entire calculus with its meta theories is formalized and proven correct in Coq. We also implement a compiler in OCaml with many supporting examples.
\end{itemize}

\section{Overall Structure}
The syntax of TLL is presented in \Cref{syntax}.

The typing rules of TLL are stratified into a level for logics and a level for programs. This is expressed formally through the two judgments depicted in \Cref{judgments}. The logical typing judgment ${\Gamma \vdash m : A}$ states that term $m$ has type $A$ under logical context $\Gamma$. The program typing judgment ${\Gamma ; \Delta \vdash m : A}$ states that term $m$ has type $A$ under logical context $\Gamma$ and program context $\Delta$.

\begin{figure}[H]
  \small
  \begin{align*}
    \Gamma \vdash m : A && \Gamma ; \Delta \vdash m : A
  \end{align*}
  \caption{Logical and Program Typing}
  \label{judgments}
\end{figure}

TLL utilizes two sorts $\Ln$ and $\Un$ to distinguish between the modalities of types (linear and non-linear respectively). When an arbitrary type $A$ is of the sort $\Ln$, we say that this type is linear. If $A$ is of sort $\Un$, we say that this type is non-linear. In order to avoid Girard's paradox, sorts can be endowed with universe levels in the usual way, but we do not do so in this paper for the sake of presentation clarity.

From a logical typing perspective, modality has no effect. This follows the intuition that hypothetical reasoning about resources does not consume them. The type system at the logical level essentially boils down to standard dependent type theory with extra modality and relevancy annotations.

At the program level, the modalities of types come into effect. Computationally relevant terms of linear types must be used exactly once and terms which are computationally irrelevant or of non-linear types can be used freely. This is accomplished by carefully controlling how contraction and weakening rules are applied to the program context $\Delta$.

For well-typed programs, a type directed erasure procedure can be carried out to remove all type annotations and sub-terms occurring in computationally irrelevant positions. The erasure soundness theorems guarantee that programs extracted by erasure always make computational progress in a manner that is compatible with their original non-erased counterparts.

\begin{figure}[H]
  \small
  \begin{tabular}{r l c l}
    sorts & $s, r, t$   &::= & $\Un$ | $\Ln$                               \\
    terms & $m,n, A, B$ &::= & $x$ | $s$                                   \\
          &             &\;| & $\PiR{t}{x : A}{B}$ | $\PiI{t}{x : A}{B}$   \\
          &             &\;| & $\lamR{t}{x : A}{m}$ | $\lamI{t}{x : A}{m}$ \\
          &             &\;| & $m\ n$ | $\square$ | ${*l}$
  \end{tabular}
  \caption{Syntax of TLL}
  \label{syntax}
\end{figure}

\section{Logical Typing}
This section describes the typing rules in the logical level. Some of these rules will appear to be redundant as type modality and computational relevancy hold no weight at the logical level where essentially everything is computationally irrelevant. However, the importance of these rules lies in their interactions with program typing which is presented in \Cref{program-typing}.

\begin{figure}[H]
  \small
  \begin{mathpar}
    \inferrule
    { }
    { \epsilon \vdash }

    \inferrule
    { \Gamma \vdash \\
      \Gamma \vdash A : s \\
      x \notin \Gamma }
    { \Gamma, x : A \vdash }
  \end{mathpar}
  \caption{Logical Context}
  \label{logical-context}
\end{figure}

\subsection{Type Formation}
\label{logical-types}
The type formation rules presented in \Cref{type-formation} appear at the logical level. They determine the canonical forms of types. An obvious departure from standard formalizations of dependent type theory is the presence of two kinds of $\Pi$-types.

The first of these is the $\PiType{0}$-type of the form $\PiI{t}{x : A}{B}$. We refer to the $\lambda$-terms inhabiting $\PiType{0}$-types at the program level as $\lamProg{0}$-programs. For $\lamProg{0}$-programs, their arguments may only be used irrelevantly within their bodies. This is similar in spirit to the $\Lambda$-quantifier of System F where type parameterized terms behave computationally the same regardless of the choice of type instantiation. However, $\PiType{0}$-types are richer than $\Lambda$-quantifiers in the sense that they can depend on arbitrary terms and not just types.

The $\PiType{1}$-type of the form $\PiR{t}{x : A}{B}$ is the usual function type. Similarly to the $\PiType{0}$-type case, we refer to the $\lambda$-terms inhabiting $\PiType{1}$-types at the program level as $\lamProg{1}$-programs. The arguments of $\lamProg{1}$-programs are allowed to be used relevantly in their bodies.

\begin{figure}[H]
  \small
  \begin{mathpar}
    \inferrule
    { \Gamma \vdash }
    { \Gamma \vdash s : \Un }

    \inferrule
    { \Gamma \vdash A : s \\
      \Gamma, x : A \vdash B : r }
    { \Gamma \vdash \PiI{t}{x : A}{B} : t }

    \inferrule
    { \Gamma \vdash A : s \\
      \Gamma, x : A \vdash B : r }
    { \Gamma \vdash \PiR{t}{x : A}{B} : t }
  \end{mathpar}
  \caption{Type Formation}
  \label{type-formation}
\end{figure}

One final detail that we want to emphasize about $\Pi$-types in TLL is the sort annotation $t$. It is clear from the typing rules here that sort $s$ of the domain, sort $r$ of the codomain and $t$ are not correlated. The $t$ annotation controls the modality of the overall $\Pi$-type intrinsically, meaning that if $t$ is set to $\Ln$, then the $\lambda$-programs inhabiting this type must be applied exactly once. In \Cref{program-typing}, we will see how the $t$ annotation imposes constraints on $\lambda$-program construction.

\subsection{Logical Terms}
The typing rules for logical terms are presented in \Cref{logical-terms}. We can see from the lack of sub-structural restrictions and the symmetry between the rules concerning $\PiType{0}$-types and $\PiType{1}$-types that the terms at the logical level are just Martin-L\"{o}f terms~\cite{martinlof} with extra annotations. The relation $\conv{A}{B}$ asserted by the last rule is the usual definitional equality relation stating that $A$ and $B$ are convertible through logical reductions (\Cref{lr-section}).

\begin{figure}[H]
  \small
  \begin{mathpar}
    \inferrule
    { \Gamma \vdash \\ x : A \in \Gamma }
    { \Gamma \vdash x : A }

    \inferrule
    { \Gamma, x : A \vdash m : B }
    { \Gamma \vdash \lamI{t}{x : A}{m} : \PiI{t}{x : A}{B} }

    \inferrule
    { \Gamma, x : A \vdash m : B }
    { \Gamma \vdash \lamR{t}{x : A}{m} : \PiR{t}{x : A}{B} }

    \inferrule
    { \Gamma \vdash m : \PiI{t}{x : A}{B} \\
      \Gamma \vdash n : A }
    { \Gamma \vdash m\ n : B[n/x] }

    \inferrule
    { \Gamma \vdash m : \PiR{t}{x : A}{B} \\
      \Gamma \vdash n : A }
    { \Gamma \vdash m\ n : B[n/x] }

    \inferrule
    { \Gamma \vdash m : A \\
      \Gamma \vdash B : s \\
      \conv{A}{B} }
    { \Gamma \vdash m : B }
  \end{mathpar}
  \caption{Logical Terms}
  \label{logical-terms}
\end{figure}

\section{Program Typing}
\label{program-typing}
The typing of programs is where our preparation at the logical level pays off. Due to the sub-structural nature of program typing, we must first understand the formation of program contexts and the constraints that can be imposed upon them before we can progress further into the presentation.

\subsection{Program Context}
A program context $\Delta$ is a sequence of triples in the form ${x \ty{s} A, y \ty{r} B, \ldots}$ where each triple is comprised of a fresh variable, a sort and a type. The well-formation of a program context $\Delta$ is defined under a logical context $\Gamma$ as the judgment ${\Gamma ; \Delta \vdash}$ whose rules are formally presented in \Cref{program-context}. Basically, for a well-formed program context according to ${\Gamma ; \Delta \vdash}$, each entry ${x \ty{s} A}$ in $\Delta$ must be correspondingly well-sorted at the logical level as ${\Gamma \vdash A : s}$. We can already see here a major role of the logical level: it provides the types that programs inhabit. 

\begin{figure}[H]
  \small
  \begin{mathpar}
    \inferrule
    { }
    { \epsilon ; \epsilon \vdash }

    \inferrule
    { \Gamma; \Delta \vdash \\
      \Gamma \vdash A : s \\
      x \notin \Gamma }
    { \Gamma, x : A; \Delta, x \ty{s} A \vdash }

    \inferrule
    { \Gamma; \Delta \vdash \\
      \Gamma \vdash A : s \\
      x \notin \Gamma }
    { \Gamma, x : A; \Delta \vdash }
  \end{mathpar}
  \caption{Program Context}
  \label{program-context}
\end{figure}

We can see from the second and third rules in \Cref{program-context} that given judgment ${\Gamma ; \Delta \vdash}$, the program context $\Delta$ forms an annotated sub-context of the logical context $\Gamma$. For the sake of readability, we implicitly assume that ${x \notin \Delta}$ whenever the notation ${\Gamma, x : A; \Delta}$ is used.

\subsection{Context Management}
Careful context management lies at the heart of sub-structural type systems. For this purpose, we introduce context merge ${\Delta_{1} \dotcup \Delta_{2}}$ and context constraint ${\Delta \triangleright s}$ whose rules are listed in \Cref{context-merge} and \Cref{context-constraint} respectively.

Context merge ${\Delta_{1} \dotcup \Delta_{2}}$ is a partial function that applies the contraction rule to overlapping $\Un$ sorted triples in program contexts $\Delta_{1}$ and $\Delta_{2}$. For $\Ln$ sorted triples, they must occur uniquely in either $\Delta_{1}$ or $\Delta_{2}$ but never both. Whenever we write ${\Delta_{1} \dotcup \Delta_{2}}$ inside typing rules, we implicitly assert that context merging is well-defined for $\Delta_{1}$ and $\Delta_{2}$.

\begin{figure}[H]
  \small
  \begin{mathpar}
    \inferrule
    { }
    { \epsilon \dotcup \epsilon = \epsilon }

    \inferrule
    { \Delta_1 \dotcup \Delta_2 = \Delta \\
      x \notin \Delta }
    { (\Delta_1, x \tU A) \dotcup (\Delta_2, x \tU A) = (\Delta, x \tU A) }
    \\

    \inferrule
    { \Delta_1 \dotcup \Delta_2 = \Delta \\
      x \notin \Delta }
    { (\Delta_1, x \tL A) \dotcup \Delta_2 = (\Delta, x \tL A) }

    \inferrule
    { \Delta_1 \dotcup \Delta_2 = \Delta \\
      x \notin \Delta }
    { \Delta_1 \dotcup (\Delta_2, x \tL A) = (\Delta, x \tL A) }
  \end{mathpar}
  \caption{Context Merge}
  \label{context-merge}
\end{figure}

For the sort indexed context constraint ${\Delta \triangleright s}$, if ${s = \Un}$ then all triples in $\Delta$ must be $\Un$ annotated. In this situation, we know from context well-formation that all types in $\Delta$ must be non-linear. On the other hand, if ${s = \Ln}$ then triples in $\Delta$ may be of both sorts. This parameterized behavior allows context constraints appearing in typing rules to work for both linear and non-linear modalities.

\begin{figure}[H]
  \small
  \begin{mathpar}
    \inferrule
    { }
    { \epsilon \triangleright s }

    \inferrule
    { \Delta \triangleright \Un }
    { \Delta, x \tU A \triangleright \Un }

    \inferrule
    { \Delta \triangleright \Ln }
    { \Delta, x \ty{s} A \triangleright \Ln }
  \end{mathpar}
  \caption{Context Constraint}
  \label{context-constraint}
\end{figure}

\subsection{General Typing}
At this point we are ready to begin the discussion on typing rules at the program level properly. An immediate difference between the program level and logical level is the lack of type formation rules at the program level. From TLL's perspective, types are hypothetical entities whose purpose is to mediate program composition. All the type formation rules for deriving the types of programs are defined at the logical level.

The first two rules at the program level are presented in \Cref{general-typing} concerning variable typing and type conversion respectively. In the variable typing rule we see that the program context $\Delta$ must contain the program variable ${x \ty{s} A}$ of interest. Furthermore, the rest of the program context is subject to constraint ${\Delta/\{x \ty{s} A\} \triangleright \Un}$ which inhibits weakening the context with variables of linear types. The conversion rule states that a program of type $A$ can be viewed as a program of type $B$ provided that $A$ and $B$ are definitionally equal and $B$ is well-sorted at the logical level.

\begin{figure}[H]
  \small
  \begin{mathpar}
    \inferrule
    { \Gamma ; \Delta \vdash \\
      x \ty{s} A \in \Delta \\
      \Delta/\{x \ty{s} A\} \triangleright \Un }
    { \Gamma ; \Delta \vdash x : A }

    \inferrule
    { \Gamma ; \Delta \vdash m : A \\
      \Gamma \vdash B : s \\
      \conv{A}{B} }
    { \Gamma ; \Delta \vdash m : B }
  \end{mathpar}
  \caption{General Program Typing}
  \label{general-typing}
\end{figure}

\subsection{Irrelevance Quantification}
In \Cref{logical-types} we introduced $\PiType{0}$-types and $\PiType{1}$-types. \Cref{irrelevance-quantification} shows how $\lamProg{0}$-programs inhabiting $\PiType{0}$-types are constructed along with their corresponding application rule.

\begin{figure}[H]
  \small
  \begin{mathpar}
    \inferrule
    { \Gamma, x : A; \Delta \vdash m : B \\
      \Delta \triangleright t }
    { \Gamma ; \Delta \vdash \lamI{t}{x : A}{m} : \PiI{t}{x : A}{B} }

    \inferrule
    { \Gamma ; \Delta \vdash m : \PiI{t}{x : A}{B} \\
      \Gamma \vdash n : A }
    { \Gamma ; \Delta \vdash m\ n : B[n/x] }
  \end{mathpar}
  \caption{Irrelevance Quantification}
  \label{irrelevance-quantification}
\end{figure}

Observe that in the premise of the $\lamProg{0}$-program construction rule, only the logical context is expanded with the parameter as ${\Gamma, x : A}$ whereas the program context $\Delta$ is left unchanged. The body of the $\lamProg{0}$-program $m$ does not have access to $x$ through the program context, so according to the program variable rule $x$ cannot be typed directly as a program in $m$. However, type annotations and irrelevant terms require only the presence of the logical context which is why $x$ can still be used irrelevantly in $m$. The side condition ${\Delta \triangleright t}$ ensures that if $\Delta$ contains linear variables, they cannot be trivially duplicated by simply packing them into a non-linear $\lamProg{0}$-program.

The application rule for $\lamProg{0}$-programs presents an interesting situation where its premise requires that $m$ be typed at the program level and $n$ be typed at the logical level. If $m$ is a $\lamProg{0}$-program, then the parameter of $m$ can only be used irrelevantly inside its body. Due to the fact that $n$ will always land in computationally irrelevant positions after $\beta$-reduction, all of $n$ is considered to be irrelevant as well.

\subsection{Relevance Quantification}
\label{relevance-typing}

The rules governing the creation and application of $\lamProg{1}$-programs are presented in \Cref{relevance-quantification}. They are similar to their irrelevance counterparts with some subtle yet important differences.

In the premise of the introduction rule for $\lamProg{1}$-programs, we see that both the logical context and program context are expanded with the parameter as ${\Gamma, x : A}$ and ${\Delta, x \ty{s} A}$ respectively. This means that $m$ can utilize $x$ both irrelevantly inside type annotations and also relevantly as a sub-program. Furthermore, if $s = \Ln$, the argument $x$ must be used exactly once inside $m$ because linearly typed variables cannot be discarded from the program context through weakening nor duplicated through contraction. In the case that $s = \Un$, the argument $x$ may be used freely inside $m$ as the structural rules are admissible on variables with non-linear types.

Since the introduction rule establishes that the arguments of $\lamProg{1}$-programs can be used relevantly inside their bodies, the application rule must account for linear resources used by the applied argument. We can see this taking place here as the argument $n$ must be typed at the program level with program context $\Delta_{2}$. Additionally, the program context $\Delta_{1}$ of $m$ is merged together with $\Delta_{2}$ as ${\Delta_{1} \dotcup \Delta_{2}}$ in the conclusion.

\begin{figure}[H]
  \small
  \begin{mathpar}
    \inferrule
    { \Gamma, x : A; \Delta, x \ty{s} A \vdash m : B \\
      \Delta \triangleright t }
    { \Gamma ; \Delta \vdash \lamR{t}{x : A}{m} : \PiR{t}{x : A}{B} }

    \inferrule
    { \Gamma ; \Delta_1 \vdash m : \PiR{t}{x : A}{B} \\
      \Gamma ; \Delta_2 \vdash n : A }
    { \Gamma ; \Delta_1 \dotcup \Delta_2 \vdash m\ n : B[n/x] }
  \end{mathpar}
  \caption{Relevance Quantification}
  \label{relevance-quantification}
\end{figure}

Careful readers may have noticed the possibility for a seemingly unsound situation to arise in the typing rule for applications. Suppose that $m$ is a $\lamProg{1}$-program whose domain is of non-linear type $A$. This means that the parameter of $m$ can be used freely inside its body. If $n$ uses linear resources in $\Delta_{2}$, then substituting $n$ into the body of $m$ could result in the duplication or leakage of resources. Unlike some prior works on linear dependent types which strictly forbid these applications from being well-typed~\cite{llf} or require multiple copies of $\Delta_{2}$~\cite{qtt}, we provide an alternative solution where applications of this form are sound using a single copy of $\Delta_{2}$. By leveraging the flexibility of TLL's two-level design, we decouple the operational semantics of the logical level from the program level and enforce a call-by-value style evaluation order in the program level semantics. When using call-by-value, $n$ must first be evaluated to a value of type $A$. This eager evaluation strategy essentially consumes all of the necessary linear resources once before $\beta$-reduction. The final value of type $A$ that substitutes into $m$ is guaranteed by the value stability theorem (\Cref{value-stability}) to be resource free which allows it to be used soundly within $m$.

\section{Program Extraction}
\label{program-extraction}

In this section, we describe the type directed procedure for erasing type annotations and irrelevant terms from TLL programs. During erasure, every term to be erased is replaced with a special constant $\square$ with no typing information nor computational behavior. We use the judgment ${\Gamma ; \Delta \vdash m \sim m' : A}$ to formally state that a program $m$ of type $A$ is erased to the extracted program $m'$. 

The following example shows the erasure of a program to a much simpler extracted form where all irrelevant terms are replaced with $\square$. Notice that the entire argument applied to the $\lamProg{0}$-program is erased in the extracted program.
\begin{align*}
  & (\lamI{\Ln}{A : \Un}{\lamR{\Ln}{x : A}{x}})\ (\PiI{\Un}{B : \Ln}{\PiR{\Un}{x : B}{B}}) \sim\\
  & (\lamI{\Ln}{A : \square}{\lamR{\Ln}{x : \square}{x}})\ \square
\end{align*}

\subsection{General Extraction}
The erasure judgment is defined in a similar fashion to program typing. We begin by presenting the erasure rules for variables and type conversion in \Cref{general-erasure}.

\begin{figure}[H]
  \small
  \begin{mathpar}
    \inferrule
    { \Gamma ; \Delta \vdash \\
      x \ty{s} A \in \Delta \\
      \Delta/\{x \ty{s} A\} \triangleright \Un }
    { \Gamma ; \Delta \vdash x \sim x : A }

    \inferrule
    { \Gamma ; \Delta \vdash m \sim m' : A \\
      \Gamma \vdash B : s \\
      \conv{A}{B} }
    { \Gamma ; \Delta \vdash m \sim m' : B }
  \end{mathpar}
  \caption{General Erasure}
  \label{general-erasure}
\end{figure}

Program variables are considered atomic by erasure in the sense that they do not contain irrelevant sub-terms. So erasure is an identity operation when applied to program variables. The type conversion erasure rule states that if a program $m$ of type $A$ can be extracted to $m'$ and $A$ is definitionally equal to some well-sorted type $B$, then $m$ can be viewed as a program of type $B$ and still be extracted to $m'$.

\subsection{Irrelevance Erasure}
The rules for performing erasure on $\lamProg{0}$-programs and their applications are presented in \Cref{irrelevance-erasure}, both of which mimic their program typing counterparts.

\begin{figure}[H]
  \small
  \begin{mathpar}
    \inferrule
    { \Gamma, x : A ; \Delta \vdash m \sim m' : B \\
      \Delta \triangleright t }
    { \Gamma ; \Delta \vdash
      \lamI{t}{x : A}{m} \sim \lamI{t}{x : \square}{m'} : \PiI{t}{x : A}{B} }

    \inferrule
    { \Gamma ; \Delta \vdash m \sim m' : \PiI{t}{x : A}{B} \\
      \Gamma \vdash n : A }
    { \Gamma ; \Delta \vdash m\ n \sim m'\ \square : B[n/x] }
  \end{mathpar}
  \caption{Irrelevance Erasure}
  \label{irrelevance-erasure}
\end{figure}

\subsection{Relevance Erasure}
Erasure for $\lamProg{1}$-programs and their applications can be carried out in an inductive manner as depicted in \Cref{relevance-erasure}. Both of these rules are straightforward as they simply push the erasure procedure structurally into their sub-programs.

\begin{figure}[H]
  \small
  \begin{mathpar}
    \inferrule
    { \Gamma, x : A ; \Delta, x \ty{s} A \vdash m \sim m' : B \\
      \Delta \triangleright t }
    { \Gamma ; \Delta \vdash
      \lamR{t}{x : A}{m} \sim \lamR{t}{x : \square}{m'} : \PiR{t}{x : A}{B} }

    \inferrule
    { \Gamma ; \Delta_1 \vdash m \sim m' : \PiR{t}{x : A}{B} \\
      \Gamma ; \Delta_2 \vdash n \sim n' : A }
    { \Gamma ; \Delta_1 \dotcup \Delta_2 \vdash m\ n \sim m'\ n' : B[n/x] }
  \end{mathpar}
  \caption{Relevance Erasure}
  \label{relevance-erasure}
\end{figure}

\section{Operational Semantics}
The exposition up until this point has been solely concerned with the static aspects of TLL such as typing and erasure. We now turn our focus to TLL's dynamic behavior by endowing it with two separate operational semantics: one for the logical level and the other for the program level. We use the relation ${m \leadsto n}$ for logical reductions and the relation ${m \Leadsto n}$ for program reductions.

\subsection{Logical Reductions}
\label{lr-section}

The reductions carried out by terms in the logical level are entirely standard. \Cref{logical-reductions} presents an excerpt of the logical reduction rules where many of the uninteresting structural cases have been elided. Unlike the reductions at the program level which are carried out using a call-by-value style evaluation order, the reductions at the logical level are not restricted to any particular evaluation order. The confluence theorem (\Cref{confluence}) for logic level reductions ensures that for any arbitrary term, every reduction strategy can ultimately be joined at a common term. Coupled with the fact that logical reductions are strongly normalizing (\Cref{strong-norm}) for well-typed logical terms, one can check the definitional equality of two terms by comparing their normal forms.

\begin{figure}[H]
  \small
  \begin{mathpar}
    \inferrule
    { m \leadsto m' }
    { m\ n \leadsto m'\ n }

    \inferrule
    { n \leadsto n' }
    { m\ n \leadsto m\ n' }
    
    \inferrule
    { }
    { (\lamI{t}{x : A}{m})\ n \leadsto m[n/x] }

    \inferrule
    { }
    { (\lamR{t}{x : A}{m})\ n \leadsto m[n/x] }
  \end{mathpar}
  \caption{Logical Reductions (Excerpt)}
  \label{logical-reductions}
\end{figure}

\subsection{Program Reductions}
As we have mentioned previously, the program level operational semantics utilizes a call-by-value style evaluation order. \Cref{program-values} lists the various value forms. We consider program variables to be values in order to allow user assumed constants to be passed around.

\begin{figure}[H]
  \small
  \begin{mathpar}
    \inferrule
    { }
    { x~\val }

    \inferrule
    { }
    { \lamI{t}{x : A}{m}~\val }

    \inferrule
    { }
    { \lamR{t}{x : A}{m}~\val }
  \end{mathpar}
  \caption{Program Values}
  \label{program-values}
\end{figure}

The reduction rules at the program level are given in \Cref{program-reduction}. Due to the fact that types and irrelevant terms are computationally inert, there are significantly fewer reduction rules at the program level than at the logical level. Among the reductions here are the $\beta_{0}$-reduction for $\lamProg{0}$-programs and $\beta_{1}$-reduction for $\lamProg{1}$-programs.

From the program level typing rules we know that arguments applied to $\lamProg{0}$-programs must be irrelevant terms. So due to its purely hypothetical nature, the irrelevant argument $n$ here will never consume actual resources. This claim is reinforced by the fact that after erasure, $n$ will be $\square$, which is completely devoid of operational behavior. Thus it is sound for the $\beta_{0}$-reduction to immediately substitute $n$ into $m$ without evaluation.

\begin{figure}[H]
  \small
  \begin{mathpar}
    \inferrule
    { m \Leadsto m' }
    { m\ n \Leadsto m'\ n }

    \inferrule
    { n \Leadsto n' }
    { m\ n \Leadsto m\ n' }

    \inferrule
    { }
    { (\lamI{t}{x : A}{m})\ n \Leadsto m[n/x] }

    \inferrule
    { v~\val }
    { (\lamR{t}{x : A}{m})\ v \Leadsto m[v/x] }
  \end{mathpar}
  \caption{Program Reductions}
  \label{program-reduction}
\end{figure}

In \Cref{relevance-typing} we have explained that the call-by-value style operational semantics at the program level allows us to assume tight resource bounds when defining the typing rules for $\lamProg{1}$-program application. This statement is realized by the $\beta_{1}$-reduction rule which requires the applied argument $v$ to be a value. Now that $v$ is a value, the resources contained within $v$ are upper bound by the value stability theorem (\Cref{value-stability}). So it is sound for the $\beta_{1}$-reduction to substitute $v$ into $m$.

\section{Meta Theory}
In this section we study the meta-theoretic properties of TLL. We organize the presentation into three subsections which are concerned with logical level theories, program level theories and program extraction theories respectively.

\subsection{Logical Theories}
The first theorem of the logical level is that of confluence. As described in \Cref{lr-section}, logical reductions do not have a fixed evaluation order so confluence is necessary to join together different reduction paths. This is especially important from an implementation perspective where definitional equality is checked by reducing terms to their normal forms. If confluence is not admissible, then certain reduction strategies may lead to a loss of valid definitional equalities. To prove the confluence theorem, we use the standard technique of showing the diamond property for parallelized logical reductions.

\begin{theorem}[Confluence of Logical Reductions]
  \label{confluence}
  If ${m \leadsto^{*} m_{1}}$ and ${m \leadsto^{*} m_{2}}$, then there exists $n$ such that ${m_{1} \leadsto^{*} n}$ and ${m_{2} \leadsto^{*} n}$.
\end{theorem}

At the logical level, types are terms inhabiting sorts. The type validity theorem shows that the types of terms are indeed valid types according to this definition. Besides substantiating the design of TLL as a dependent type system, the type validity theorem also provides a great deal of utility when proving other theorems as it allows types to be viewed as terms. This enables various inversion lemmas to be applicable to types.

\begin{theorem}[Type Validity]
  For any logical typing ${\Gamma \vdash m : A}$, there exists sort $s$ such that ${\Gamma \vdash A : s}$ is derivable.
\end{theorem}

The sort of a TLL type determines its modality. A type inhabiting the $\Ln$ sort is linear and a type inhabiting the $\Un$ sort is non-linear. With the sorts of types playing such a crucial role in the sub-structural type system at the program level, it is important to show that no ambiguity arises when assigning sorts to types at the logical level. The sort uniqueness theorem states that the sort of a particular type is unique thus preventing contradictory situations where a type is both $\Ln$ sorted and $\Un$ sorted. When viewed in conjunction with type validity, these theorems show that there always exists a unique sort for the type of a term to inhabit.

\begin{theorem}[Sort Uniqueness] \label{sort-unique}
  If there are logical typings ${\Gamma \vdash A : s}$ and ${\Gamma \vdash A : t}$, then $s = t$.
\end{theorem}

The standard subject reduction theorem is admissible for well-typed logical terms. This means that the types of logical terms are preserved by reductions. Properties and theorems that are derived from the logical typing judgment can be propagated across reductions as well. Furthermore, subject reduction also ensures that a reduction based definitional equality checker never alters the types of its candidates.

\begin{theorem}[Logical Subject Reduction]
  If there are logical typing ${\Gamma \vdash m : A}$ and reduction ${m \leadsto n}$, then ${\Gamma \vdash n : A}$ is derivable.
\end{theorem}

Finally, we have the strong normalization theorem at the logical level. Assuming that sorts are always implicitly labeled with universe levels in the usual way (i.e., $s_{l} : \Un_{l+1}$), then universe inconsistencies can be ruled out by the type system. At this point, the logical level of TLL can be modeled in Martin-L\"{o}f type theory (MLTT)~\cite{martinlof} in a straightforward manner that preserves its reduction behavior. The rules for carrying out the modeling procedure are given formally in \Cref{mltt-model}. Basically, an MLTT model of logical TLL collapses the two sorts into one and inductively strips terms of their modality and relevancy annotations.

\begin{figure}[H]
  \small
  \begin{align*}
    \model{x}                  &= x \\
    \model{\Un}                &= \text{Type} \\
    \model{\Ln}                &= \text{Type} \\
    \model{\PiI{t}{x : A}{B}}  &= \Pi(x : \model{A}).\model{B} \\ 
    \model{\PiR{t}{x : A}{B}}  &= \Pi(x : \model{A}).\model{B} \\ 
    \model{\lamI{t}{x : A}{B}} &= \lambda(x : \model{A}).\model{B} \\ 
    \model{\lamR{t}{x : A}{B}} &= \lambda(x : \model{A}).\model{B} \\ 
    \model{m\ n}               &= \model{m}\ \model{n}
  \end{align*}
  \caption{Logical TLL in Martin-L\"{o}f}
  \label{mltt-model}
\end{figure}

After naturally extending the modeling procedure to all types appearing in TLL logical contexts, the following two lemmas can be proven. These results show that this model is indeed sound with regards to logical reduction. By virtue of the strong normalization property for MLTT~\cite{Coquand06}, TLL must be strongly normalizing as well.

\begin{lemma}[Logical Type Model]
  Given a TLL logical typing judgment ${\Gamma \vdash_{\text{TLL}} m : A}$, the judgment ${\model{\Gamma} \vdash_{\text{MLTT}} \model{m} : \model{A}}$ can be derived in Martin-L\"{o}f type theory.
\end{lemma}

\begin{lemma}[Logical Reduction Model]
  Given a TLL logical reduction ${m \leadsto_{\text{TLL}} n}$, the reduction ${\model{m} \leadsto_{\text{MLTT}} \model{n}}$ can be derived in Martin-L\"{o}f type theory.
\end{lemma}

\begin{theorem}[Logical Strong Normalization]
  \label{strong-norm}
  For any TLL term $m$ with logical typing ${\Gamma \vdash m : A}$, it is strongly normalizing.
\end{theorem}

\subsection{Program Theories}
TLL prohibits weakening the program context with variables of linear types to prevent the discarding of resources. However, the logical context can be independently weakened by itself since hypothetical resources can always be assumed freely. Moreover, weakening is admissible for the program context if the weakened variable is of non-linear type. These observations are expressed formally in the following pair of weakening lemmas.

\begin{lemma}[Program 0-Weakening]
  For valid program typing ${\Gamma ; \Delta \vdash m : A}$ and logical typing ${\Gamma \vdash B : s}$, the judgment ${\Gamma, x : B ; \Delta \vdash m : A}$ is derivable for any $x \notin \Gamma$.
\end{lemma}

\begin{lemma}[Program 1-Weakening]
  For valid program typing ${\Gamma ; \Delta \vdash m : A}$ and logical typing ${\Gamma \vdash B : \Un}$, the judgment ${\Gamma, x : B ; \Delta, x \tU B \vdash m : A}$ is derivable for any $x \notin \Gamma$.
\end{lemma}

A common drawback of stratified type systems is the lack of code sharing between language fragments. Structures performing similar tasks must be implemented independently in the different layers of the language. Libraries with large amounts of redundant code can become difficult to scale and maintain so it is important to reduce code duplication to the best of our abilities. The program reflection theorem tackles the code redundancy problem by allowing us to freely reflect well-typed programs into the logical level. This essentially allows the sharing of all code written in the program level with the logical level.

\begin{theorem}[Program Reflection]
  For any program typing ${\Gamma ; \Delta \vdash m : A}$, logical typing ${\Gamma \vdash m : A}$ is derivable.
\end{theorem}

Arbitrary TLL programs may utilize linear resources to compute a final non-linear value. So despite these programs being of non-linear types, they cannot be freely duplicated without breaking the no-contraction principle. However for programs in value form, the value stability theorem gives an upper bound on the resources they are allowed to consume. For a linearly typed value, it will always be used exactly once and consequently any resource held by the value is used exactly once as well. For values of non-linear type, the context constraint ${\Delta \triangleright \Un}$ prevents resources from occurring inside the value which allows it to be duplicated soundly.

\begin{theorem}[Value Stability]
  \label{value-stability}
  If there is value $v$ with program typing ${\Gamma ; \Delta \vdash v : A}$ and ${\Gamma \vdash A : s}$, then ${\Delta \triangleright s}$.
\end{theorem}

The program level supports its own version of the subject reduction theorem that is defined on the program typing judgment and program reductions. Although there are less typing rules at the program level when compared to the logical level, the program subject reduction is more difficult to prove than its logical counterpart. This is due to the necessity of carefully tracking changes in the program context during variable substitution. The two following lemmas describe the interactions between substitution and contexts.

\begin{lemma}[Program 0-Substitution]
  If there are program typing ${\Gamma, x : A; \Delta \vdash m : B}$ and logical typing ${\Gamma \vdash n : A}$, then ${\Gamma ; \Delta \vdash m[n/x] : B[n/x]}$ is derivable.
\end{lemma}

\begin{lemma}[Program 1-Substitution]
  If there are program typings ${\Gamma, x : A; \Delta_{1}, x \ty{s} A \vdash m : B}$ and ${\Gamma ; \Delta_{2} \vdash n : A}$ and context constraint ${\Delta_{2} \triangleright s}$, then ${\Gamma ; \Delta_{1} \dotcup \Delta_{2} \vdash m[n/x] : B[n/x]}$ is derivable.
\end{lemma}

\begin{theorem}[Program Subject Reduction]
  For any program typing ${\epsilon ; \epsilon \vdash m : A}$ and reduction ${m \Leadsto n}$, there is ${\epsilon ; \epsilon \vdash n : A}$.
\end{theorem}

To show that well-typed programs ``cannot go wrong'', we prove the following progress theorem. When viewed together with the program subject reduction theorem, it is clear that closed TLL programs will not get stuck during evaluation.

\begin{theorem}[Program Progress]
  If there is program typing ${\epsilon ; \epsilon \vdash m : A}$, then $m$ is either a value or there exists $n$ such that ${m \Leadsto n}$.
\end{theorem}

\subsection{Program Extraction}
Introduced in \Cref{program-extraction}, the program erasure procedure is carried out by inductively erasing type annotations and irrelevant terms occurring inside programs. The erasure existence theorem shows that extraction is well-defined for all well-typed TLL programs.

\begin{theorem}[Erasure Existence]
  For any well typed program ${\Gamma ; \Delta \vdash m : A}$, there exists an extracted version of it $m'$ such that the erasure relation ${\Gamma ; \Delta \vdash m \sim m' : A}$ is derivable.
\end{theorem}

After erasure has been successfully carried out on well-type programs, the extracted results retain only the relevant parts of the original program. It is now important to show that these extracted programs still behave as expected of their original selves computationally. We accomplish this by proving instrumented subject reduction and progress theorems.

The first theorem to establish the connection between original programs and their extracted forms is the erasure subject reduction theorem. Erasure subject reduction tells us that if a program reduction ${m' \Leadsto n'}$ can be triggered for the extracted $m'$ of a well-typed program $m$, then there exists a well-type program $n$ that extracts to $n'$ and the reduction ${m \Leadsto n}$ exists on the original forms.

\begin{theorem}[Erasure Subject Reduction]
  For any erasure relation ${\epsilon ; \epsilon \vdash m \sim m' : A}$ and reduction ${m' \Leadsto n'}$, there exists program $n$ such that the following diagram commutes.
  \begin{center}
    \begin{tikzcd}
      {\epsilon ; \epsilon} &[-18pt] m &[-7pt] {m'} &[-20pt] A \\
      {\epsilon ; \epsilon} &[-18pt] n &[-7pt] {n'} &[-20pt] A
      \arrow["\sim"{description}, draw=none, from=1-2, to=1-3]
      \arrow["\vdash"{description}, draw=none, from=1-1, to=1-2]
      \arrow["{:}"{description}, draw=none, from=1-3, to=1-4]
      \arrow["\Leadsto"{marking}, draw=none, from=1-2, to=2-2]
      \arrow["\Leadsto"{marking}, draw=none, from=1-3, to=2-3]
      \arrow["\vdash"{description}, draw=none, from=2-1, to=2-2]
      \arrow["{:}"{description}, draw=none, from=2-3, to=2-4]
      \arrow["\sim"{description}, draw=none, from=2-2, to=2-3]
    \end{tikzcd}
  \end{center}
\end{theorem}

Through the erasure progress theorem we show that the extracted forms of well-typed programs exhibit the same property of never getting stuck during evaluation.

\begin{theorem}[Erasure Progress]
  \label{erasure-progress}
  If there is erasure relation ${\epsilon ; \epsilon \vdash m \sim m' : A}$, then $m'$ is a value or there exists $n'$ such that $m' \Leadsto n'$.
\end{theorem}

\section{Heap Semantics}
In order to realize the abstract notion of linear resource, we develop a heap semantics in the style of Turner and Wadler~\cite{turner99} where heap memory serves as resource to the language runtime. At the heart of the semantics is the following heap indexed reduction relation.
\begin{align*}
  H_{1} ; m_{1} \Leadsto H_{2} ; m_{2}
\end{align*}
The term $m_{1}$ here is an extracted program that may contain pointers to cells in heap $H_{1}$ storing values. As $m_{1}$ evaluates, new cells in the heap are allocated to store intermediate values that are generated. The effects of a single evaluation step are reflected in the updated heap $H_{2}$ and program $m_{2}$. If evaluating $m_{1}$ requires dereferencing pointers, then depending on the modality of the pointed to cell, the heap is either unchanged for non-linear cells or deallocated for linear cells. Such a semantics closely models low level implementations of functional languages where heap memory is dynamically allocated to store closures and other various data structures. Through this heap semantics we show that linearity enforces safe memory usage and that linearly typed programs are guaranteed to run memory clean without the need for runtime garbage collection.

\subsection{Heaps}
Heaps $H$ are maps from unique locations $l$ to sort annotated values. Generally, a heap is of the following form:
\begin{align*}
  H ::= \{l_{1} \mapsto_{s_{1}} v_{1}, l_{2} \mapsto_{s_{2}} v_{2}, \ldots, l_{k} \mapsto_{s_{k}} v_{k}\}
\end{align*}
Each entry ${l_{i} \mapsto_{s_{i}} v_{i}}$ denotes a mapping from location $l_{i}$ to value $v_{i}$ of $s_{i}$ modality. In particular, if $s_{i} = \Un$ then looking up location $l_{i}$ in the heap will not cause any changes. However, if $s_{i} = \Ln$ then looking up location $l_{i}$ in the heap will remove the mapping from the heap. Formally, we introduce the relation $\lookup{H_{1}}{l}{v}{H_{2}}$ with rules presented in \Cref{lookup}, stating that looking up location $l$ in heap $H_{1}$ results in value $v$ and heap $H_{2}$. Depending on the mapping modality between $l$ and $m$, the resulting heap $H_{2}$ after lookup is equal to either $H_{1}$ in the $\Un$ case or $H_{1}/\{l \mapsto_{\Ln} v\}$ in the $\Ln$ case.

\begin{figure}[H]
  \small
  \begin{mathpar}
    \inferrule
    { (l \mapsto_{\Un} v) \in H }
    { \lookup{H}{l}{v}{H} }

    \inferrule
    { (l \mapsto_{\Ln} v) \in H }
    { \lookup{H}{l}{v}{H/\{l \mapsto_{\Ln} v\}} }
  \end{mathpar}
  \caption{Heap Lookup}
  \label{lookup}
\end{figure}

Notice that the entries ${l \mapsto_{s} v}$ of heaps are similar to the triples ${x \ty{s} A}$ of program contexts. While one maps locations to values, the other maps variables to types. Both kinds of mappings are annotated by sort $s$. Taking advantage of these commonalities, we overload the merge operator ${H_{1} \dotcup H_{2}}$ and constraint ${H \triangleright s}$ to work for heaps as well. So instead of operating over variables in contexts, they operate over locations in heaps.

\subsection{Heap Reductions}
The rules of heap reductions are presented in \Cref{heap-reductions}. These rules essentially form a modified version of the program level call-by-value semantics. When a program is evaluated to value $v$, a memory cell at a fresh location $l$ in the heap is allocated to store $v$. For example, since $\lambda$-programs are considered values, a fresh cell with the same modality as the $\lambda$-program is allocated to store it. Now if a pointer expression ${*l}$ is encountered, we know that it points to a value located in the heap. The application rules utilize this fact to enforce the call-by-value evaluation order for relevant applications.

\begin{figure}[H]
  \small
  \begin{mathpar}
    \inferrule
    { l \notin H }
    { H ; \lamI{t}{x : \square}{m} \Leadsto
      H \cup \{l \mapsto_t \lamI{t}{x : \square}{m}\} ; {*l} }

    \inferrule
    { l \notin H }
    { H ; \lamR{t}{x : \square}{m} \Leadsto
      H \cup \{l \mapsto_t \lamR{t}{x : \square}{m}\} ; {*l} }

    \inferrule
    { H ; m \Leadsto H' ; m' }
    { H ; m\ n \Leadsto H' ; m'\ n }

    \inferrule
    { H ; n \Leadsto H' ; n' }
    { H ; m\ n \Leadsto H' ; m\ n' }\\

    \inferrule
    { \lookup{H}{l}{\lamI{t}{x : \square}{m}}{H'} }
    { H ; {*l}\ n \Leadsto H' ; m[\square/x] }

    \inferrule
    { \lookup{H}{l_1}{\lamR{t}{x : \square}{m}}{H'} }
    { H ; {*l_1}\ {*l_2} \Leadsto m[{*l_2}/x] }
  \end{mathpar}
  \caption{Heap Reductions}
  \label{heap-reductions}
\end{figure}

Generally for redexes in the heap semantics, pointers are expected in place of values in the standard program semantics. For example, the application of two pointers ${*l_{1}}\ {*l_{2}}$ is considered a $\beta_{1}$-redex if ${*l_{1}}$ points to a relevant $\lamProg{1}$-program in the heap. To reduce this redex, the argument pointer ${*l_{2}}$ is substituted into the body of the $\lamProg{1}$-program referenced by ${*l_{1}}$. Likewise, the application form ${*l}\ n$ is considered a $\beta_{0}$-redex if ${*l}$ points to an irrelevant $\lamProg{0}$-program in the heap. For extracted programs satisfying the erasure relation, the argument $n$ here must be $\square$ so we immediately substitute $\square$ into the body of the $\lamProg{0}$-program referenced by ${*l}$.

\subsection{Pointer Resolution}
We do not extend our typing rules or standard operational semantics to cover pointer expressions, so whenever there is ${\Gamma ; \Delta \vdash m : A}$ or ${m \Leadsto n}$ we know that all terms involved must not contain pointers. Instead, we introduce a new judgment ${H ; m \sim n}$ presented in \Cref{pointer-resolution} which recursively dereferences all pointers in $m$ using heap $H$ until there are no pointers occurring in $n$.

\begin{figure}[H]
  \small
  \begin{mathpar}
    \inferrule
    { H \triangleright \Un }
    { H ; x \sim x }

    \inferrule
    { H \triangleright t \\
      H ; m' \sim m }
    { H ; \lamI{t}{x : \square}{m'} \sim \lamI{t}{x : \square}{m} }

    \inferrule
    { H \triangleright t \\
      H ; m' \sim m }
    { H ; \lamR{t}{x : \square}{m'} \sim \lamR{t}{x : \square}{m} }

    \inferrule
    { H ; m' \sim m }
    { H ; m'\ \square \sim m\ \square }

    \inferrule
    { H_1 ; m' \sim m \\
      H_2 ; n' \sim n }
    { H_1 \dotcup H_2 ; m'\ n' \sim m\ n }

    \inferrule
    { \lookup{H}{l}{v'}{H'} \\
      H' ; v' \sim v }
    { H ; {*l} \sim v }
  \end{mathpar}
  \caption{Pointer Resolution}
  \label{pointer-resolution}
\end{figure}

When defining the judgment ${H ; m \sim n}$, care is taken to ensure that dereferencing pointers obey the no-weakening and no-contraction principles analogously to the program typing rules. In other words, pointers to linear mappings in $H$ are dereferenced only once. This is accomplished by enforcing side conditions ${H_{1} \dotcup H_{2}}$ and $H \triangleright s$ which basically perform the same roles as their typing counterparts. The only rule without a typing counterpart is the last rule where it initially dereferences pointer ${*l}$ to value $v'$ and recursively resolves all pointers in $v'$ using the remaining heap $H'$.

To reintroduce typing information back into programs containing pointer expressions, we develop the ternary relation of well-resolved programs presented in \Cref{well-resolved} that unites the erasure relation of \Cref{program-extraction} and pointer resolution. For a valid instance of well-resolved ${H \vdash a \sim b \sim c : A}$ we know that $a$ is a well-type program that extracts to $b$ and resolving the pointers in $c$ also gives us $b$. Essentially, the extracted program $b$ serves to bridge the well-typed original program $a$ and the pointer program $c$.

\begin{figure}[H]
  \small
  \begin{mathpar}
    \inferrule
    { \epsilon ; \epsilon \vdash a \sim b : A \\
      H ; c \sim b }
    { H \vdash a \sim b \sim c : A }
  \end{mathpar}
  \caption{Well-Resolved}
  \label{well-resolved}
\end{figure}

From the definition of pointer resolution it is clear that the contents in heap $H$ greatly influence the outcome of resolving pointer program $m$ in judgment ${H ; m \sim n}$. To characterize the heaps produced during heap reduction, we introduce in \Cref{wr-heaps} a judgment $H~\wrheap$ stating that $H$ is of a form that could plausibly be produced by heap reduction. Basically, all locations in $H$ map to closed values. This means that performing lookup in heaps satisfying the $\wrheap$ property is guaranteed to be productive because a value is retrieved after one level of indirection. The $\wrheap$ property serves as an inductive invariant in the proof of heap semantics soundness, informing us of the structure of heaps at every reduction step.

\begin{figure}[H]
  \small
  \begin{mathpar}
    \inferrule
    { }
    { \epsilon~\wrheap }

    \inferrule
    { \FV{m}/\{x\} = \emptyset \\
      H~\wrheap \\
      l \notin H }
    { H \cup \{l \mapsto_t \lamI{t}{x : \square}{m}\}~\wrheap }

    \inferrule
    { \FV{m}/\{x\} = \emptyset \\
      H~\wrheap \\
      l \notin H }
    { H \cup \{l \mapsto_t \lamR{t}{x : \square}{m}\}~\wrheap }
  \end{mathpar}
  \caption{WR-Heaps}
  \label{wr-heaps}
\end{figure}

\subsection{Soundness of Heap Semantics}
The soundness of heap semantics is again justified through progress-preservation style theorems. Due to the fact that these theorems must now account for program typing, erasure, pointer resolution, heap reductions and many other concepts simultaneously, their statements and proofs become significantly more involved than previous versions.

The first theorem that we present is resolution stability. It is reminiscent of the value stability theorem (\Cref{value-stability}). But instead of exploring the constraints set by the modality of values on their program contexts as in the value stability case, the resolution stability theorem derives constraints on the mappings inside heaps.

\begin{theorem}[Resolution Stability]
  Given valid instances of well-resolved ${H \vdash a \sim b \sim c : A}$, logical typing ${\epsilon \vdash A : s}$ and ${H~\wrheap}$, if $b$ is a value then heap $H$ can be upper bound by constraint ${H \triangleright s}$.
\end{theorem}

The heap subject reduction theorem propagates the well-resolved and $\wrheap$ invariants across heap reduction, ensuring that pointer programs well-resolved in wr-heaps always reduce to programs that are well-resolved in wr-heaps. Additionally, heap reductions agree with iterated steps in the standard semantics for original programs and extracted programs.

\begin{theorem}[Heap Subject Reduction]
  Given instances of well-resolved ${H \vdash a \sim b \sim c : A}$ and ${H~\wrheap}$, then for heap reduction ${H ; c \Leadsto H' ; c'}$ there exist $a'$ and $b'$ such that the following judgments ${H' \vdash a' \sim b' \sim c' : A}$, ${H'~\wrheap}$, ${a \Leadsto^{*} a'}$ and ${b \Leadsto^{*} b'}$ all hold.
\end{theorem}

Finally, the heap progress theorem shows that for any pointer program $c$ that is well-resolved in a heap $H$ satisfying the $\wrheap$ condition, either there exists a heap reduction ${H ; c \Leadsto H' ; c'}$ or $c$ is a pointer. From the definition of $\wrheap$ we know that all elements contained in the heap are values. In the case that $c$ is a pointer, dereferencing $c$ yields a value.

\begin{theorem}[Heap Progress]
  Given valid instances of well-resolved ${H \vdash a \sim b \sim c : A}$ and ${H~\wrheap}$, then either there exist heap $H'$ and program $c'$ such that there is reduction ${H ; c \Leadsto H' ; c'}$ or there exists a location $l$ such that ${c = {*l}}$.
\end{theorem}

Starting from a well-typed closed program $m$ and an empty heap, due to the fact that empty heaps are degenerate wr-heaps and well-typed closed programs are trivially well-resolved in empty heaps, the heap subject reduction theorem and heap progress theorem allow for heap reductions to be repeatedly generated and applied until a value referencing pointer is reached. The resolution stability theorem tells us that the heap at this point can be constrained by ${H \triangleright s}$ where $s$ is the sort of the original program $m$'s type. In practice, if the designated starting \prog{main} expression is required to be of a non-linear type, all allocated heap memory will be safely freed by the time that the program terminates.

\section{Extensions}
In this section, we describe some useful extensions to the core TLL language for program development and reasoning. The meta theoretic results presented in the previous sections can all be extended naturally to cover these additions. In fact, our theorems are all proven assuming the inclusion of these extensions.

\subsection{Propositional Equality}
\label{prop-equality}
The first extension that we present is propositional equality with logical typing rules given in \Cref{logical-prop-equality}. This is the usual propositional equality found in intensional dependent type theories which allows one to posit equality between two terms. Equality and its proofs exist purely for reasoning purposes so they can only be derived at the logical level. However, a proof of equality constructed at the logical level can be eliminated at the program level using the rule shown in \Cref{program-prop-equality}. This allows the type-casting of programs with a proof that the target type is propositionally equal to the original type. A simple example of program level equality elimination in practice is the type-casting of a length indexed vector $\textit{ls}$ with type $\textit{vec}\ (x + y)\ A$ to type $\textit{vec}\ (y + x)\ A$ by appealing to the proof that addition is commutative.

\begin{figure}[H]
  \small
  \begin{mathpar}
    \inferrule
    { \Gamma \vdash A : s \\
      \Gamma \vdash m : A \\
      \Gamma \vdash n : A }
    { \Gamma \vdash \iden{A}{m}{n} : \Un }

    \inferrule
    { \Gamma \vdash m : A }
    { \Gamma \vdash \refl{m} : \iden{A}{m}{m} }

    \inferrule
    { \Gamma, x : A, p : \iden{A}{m}{x} \vdash B : s \\
      \Gamma \vdash H : B[m/x,\refl{m}/p] \\
      \Gamma \vdash P : \iden{A}{m}{n} }
    { \Gamma \vdash \idenElim{[x,p]B}{H}{P} : B[n/x,P/p] }
  \end{mathpar}
  \caption{Propositional Equality (Logical Rules)}
  \label{logical-prop-equality}
\end{figure}
\vspace{-1em}
\begin{figure}[H]
  \small
  \begin{mathpar}
    \inferrule
    { \Gamma, x : A, p : \iden{A}{m}{x} \vdash B : s \\
      \Gamma ; \Delta \vdash H : B[m/x,\refl{m}/p] \\
      \Gamma \vdash P : \iden{A}{m}{n} }
    { \Gamma ; \Delta \vdash \idenElim{[x,p]B}{H}{P} : B[n/x,P/p] }
  \end{mathpar}
  \caption{Propositional Equality (Program Rules)}
  \label{program-prop-equality}
\end{figure}

The most interesting aspect of TLL propositional equality lies in its different reduction behaviors at the logical level and at the program level, the rules of which are presented in \Cref{program-prop-reduction}. Notice that in the logical reduction rule ${\idenElim{[x,p]A}{H}{\refl{m}} \leadsto H}$, the proof of equality must be of the form $\refl{m}$ in order to trigger reduction. This is to ensure that $\idenElim{[x,p]A}{H}{\refl{m}}$ and $H$ have definitionally equal types. This seems to indicate that propositional equality is computationally relevant in the sense that an equality eliminator $\idenElim{[x,p]A}{H}{P}$ occurring at the program level ought to carry around proof $P$ and reduce it to $\refl{m}$. From the actual program level reduction rule $\idenElim{[x,p]A}{H}{P} \Leadsto H$ we can see that this first impression is misleading: the program level equality eliminator does not impose any restrictions on $P$ and immediately reduces to $H$. The soundness of this rule is due to the fact that program reductions are not performed under context. By the time an equality eliminator is evaluated, logical strong normalization and canonicity guarantee the existence of a $\refl{m}$ proof that $P$ is logically reducible to. In other words, equality eliminators at the program level reduce their proofs \textit{conceptually} but not literally.

\begin{figure}[H]
  \small
  \begin{align*}
    \idenElim{[x,p]A}{H}{\refl{m}} \leadsto H &&
    \idenElim{[x,p]A}{H}{P} \Leadsto H
  \end{align*}
  \caption{Propositional Equality Reduction}
  \label{program-prop-reduction}
\end{figure}

After recognizing that equality proofs are computationally irrelevant at the program level, we define the erasure procedure for equality eliminators in \Cref{erasure-prop-equality} that removes the equality proof $P$ entirely. These erased programs can be represented and evaluated much more efficiently at runtime than their compile time logical counterparts.

\begin{figure}[H]
  \small
  \begin{mathpar}
    \inferrule
    { \Gamma, x : A, p : \iden{A}{m}{x} \vdash B : s \\
      \Gamma ; \Delta \vdash H \sim H' : B[m/x,\refl{m}/p] \\
      \Gamma \vdash P : \iden{A}{m}{n} }
    { \Gamma ; \Delta \vdash \idenElim{[x,p]B}{H}{P} \sim \idenElim{\square}{H'}{\square} : B[n/x,P/p] }
  \end{mathpar}
  \caption{Propositional Equality (Erasure Rules)}
  \label{erasure-prop-equality}
\end{figure}

\subsection{Subset Pairs}
\label{subset-pairs}
Leveraging the distinction between proofs and programs in TLL, we encode a variation of $\Sigma$-types often referred to as subset types in the verification community. For a subset type of the form $\SigI{t}{x : A}{B}$, its canonical inhabitant will be a pair of the form $\pairI{m}{n}{t}$ where $m$ is a relevant payload of type $A$ and $n$ is an irrelevant proof of the dependent type $B$. A common use case for subset types is to refine programs by the properties they satisfy, essentially carving out a \textit{subset} of the original type.

The logical rules for subset types are presented in \Cref{logical-subset-pair} and the program rules are presented in \Cref{program-subset-pair}. The side condition ${(t = \Un) \Rightarrow (s = \Un)}$ is required by the subset type formation rule to prevent resource leakage via packing linear payloads into non-linear subset pairs. Notice how in the program rule for subset pair construction the payload $m$ is typed at the program level whereas the term $n$ is typed at the logical level. This indicates that $m$ is computationally relevant and $n$ is computationally irrelevant. From an erasure perspective, a program of the form $\pairI{m}{n}{t}$ is erased to $\pairI{m}{\square}{t}$.

\begin{figure}[H]
  \small
  \begin{mathpar}
    \inferrule
    { \Gamma \vdash A : s \\
      \Gamma, x : A \vdash B : r \\
      (t = \Un) \Rightarrow (s = \Un) }
    { \Gamma \vdash \SigI{t}{x : A}{B} : t }

    \inferrule
    { \Gamma \vdash \SigI{t}{x : A}{B} : t \\
      \Gamma \vdash m : A \\
      \Gamma \vdash n : B[m/x] }
    { \Gamma \vdash \pairI{m}{n}{t} : \SigI{t}{x : A}{B} }

    \inferrule
    { \Gamma, z : \SigI{t}{x : A}{B} \vdash C : s \\
      \Gamma \vdash m : \SigI{t}{x : A}{B} \\
      \Gamma, x : A, y : B \vdash n : C[\pairI{x}{y}{t}/z] }
    { \Gamma \vdash \SigElim{[z]C}{m}{[x,y]n} : C[m/z] }
  \end{mathpar}
  \caption{Subset Pairs (Logical Rules)}
  \label{logical-subset-pair}
\end{figure}
\vspace{-1em}
\begin{figure}[H]
  \small
  \begin{mathpar}
    \inferrule
    { \Gamma \vdash \SigI{t}{x : A}{B} : t \\
      \Gamma ; \Delta \vdash m : A \\
      \Gamma \vdash n : B[m/x] }
    { \Gamma ; \Delta \vdash \pairI{m}{n}{t} : \SigI{t}{x : A}{B} }

    \inferrule
    { \Gamma, z : \SigI{t}{x : A}{B} \vdash C : s \\
      \Gamma ; \Delta_1 \vdash m : \SigI{t}{x : A}{B} \\
      \Gamma, x : A, y : B; \Delta_2, x \ty{r} A \vdash n : C[\pairI{x}{y}{t}/z] }
    { \Gamma ; \Delta_1 \dotcup \Delta_2 \vdash \SigElim{[z]C}{m}{[x,y]n} : C[m/z] }
  \end{mathpar}
  \caption{Subset Pairs (Program Rules)}
  \label{program-subset-pair}
\end{figure}

The following example shows applying erasure to a pair of type ${\SigI{\Un}{x : \nat}{x + 1 =_{\nat} 2}}$. Notice that the proof component of the pair $(\refl 2)$ is completely removed by erasure. These subset pairs realize the principle of computational irrelevancy for program properties.
\begin{align*}
  \vdash \pairI{1}{\refl{2}}{\Un} \sim \pairI{1}{\square}{\Un} : \SigI{\Un}{x : \nat}{x + 1 =_{\nat} 2}
\end{align*}

Standard dependent pairs where both components are computationally relevant can be defined in a straightforward manner. The typing rules and semantics for relevant pairs are fully formalized in our Coq development, but for the sake of saving space we do not present them here.

\subsection{Additive Pairs}
To integrate the additive fragment of Linear Logic~\cite{girard} into TLL, we introduce $\&$-types as an extension. The logical rules are presented in \Cref{logical-additive-pair} and the program rules are presented in \Cref{program-additive-pair}. Intuitively, a $\&$-type of the form $\with{A}{B}{t}$ represents the pairing of two delayed computations of types $A$ and $B$ respectively. Canonical inhabitants of $\with{A}{B}{t}$ are additive pairs of the form $\apair{m}{n}{t}$.

Of the rules depicted here, the most interesting is the program typing rule governing the construction of additive pairs. Notice that in the premise, both components $m$ and $n$ are typed in the same program context $\Delta$. Furthermore, the conclusion only assumes a single copy of $\Delta$. This is a realization of the additive fragment of Linear Logic~\cite{girard}. Due to the fact that $m$ and $n$ are delayed computations, only one of the two will ultimately be projected out and evaluated. So only a single copy of $\Delta$ is committed to the component that actually gets evaluated.

\begin{figure}[H]
  \small
  \begin{mathpar}
    \inferrule
    { \Gamma \vdash A : s \\
      \Gamma \vdash B : r }
    { \Gamma \vdash \with{A}{B}{t} : t }

    \inferrule
    { \Gamma \vdash m : A \\
      \Gamma \vdash n : B }
    { \Gamma \vdash \apair{m}{n}{t} : \with{A}{B}{t} }

    \inferrule
    { \Gamma \vdash m : \with{A}{B}{t} }
    { \Gamma \vdash \projL{m} : A }

    \inferrule
    { \Gamma \vdash m : \with{A}{B}{t} }
    { \Gamma \vdash \projR{m} : B }
  \end{mathpar}
  \caption{Additive Pairs (Logical Rules)}
  \label{logical-additive-pair}
\end{figure}
\vspace{-1em}
\begin{figure}[H]
  \small
  \begin{mathpar}
    \inferrule
    { \Gamma ; \Delta \vdash m : A \\
      \Gamma ; \Delta \vdash n : B \\
      \Delta \triangleright t }
    { \Gamma ; \Delta \vdash \apair{m}{n}{t} : \with{A}{B}{t} }

    \inferrule
    { \Gamma ; \Delta \vdash m : \with{A}{B}{t} }
    { \Gamma ; \Delta \vdash \projL{m} : A }

    \inferrule
    { \Gamma ; \Delta \vdash m : \with{A}{B}{t} }
    { \Gamma ; \Delta \vdash \projR{m} : B }
  \end{mathpar}
  \caption{Additive Pairs (Program Rules)}
  \label{program-additive-pair}
\end{figure}

\section{Implementation and Application}
In this section, we describe language features implemented in the TLL compiler with simple examples on how they can be used to effectively construct and verify programs.


\subsection{Linear Inductive Types}
\label{inductive}
The TLL compiler supports user defined inductive types in the style of CIC~\cite{cic}. \Cref{linear-list} demonstrates how a linear list can be defined. In this definition, the \textit{arity} of the \prog{llist} type constructor ends in sort \prog{L}. Once \prog{llist} is fully applied to an arbitrary linear type \prog{A}, the resulting type \prog{llist A} will be of sort \prog{L} which requires that the lists inhabiting this type are used exactly once. So basically, by varying the sorts of type arities and constructor arguments, we can define different combinations of linear and non-linear inductive types.

\begin{figure}[H]
  \scriptsize
  \begin{tllisting}
inductive llist (A : L) : L =
| lnil
| lcons of (hd : A) (tl : llist A)
  \end{tllisting}
  \caption{Linear Lists}
  \label{linear-list}
\end{figure}

\Cref{lappend} defines a function~\footnote{Underscores can be used as implicit arguments which are inferred through unification.} for appending two linear lists. The usage of the toplevel keyword \prog{program} here allows the \prog{lappend} function to be applicable at the program level. Since types can only appear in computationally irrelevant positions at the program level, the type parameter \prog{A} is quantified irrelevantly as a $\lamProg{0}$ argument. Additionally, the body of \prog{lappend} is subject to linear type-checking because it can be used at the program level. The C code emitted for \prog{lappend} will reclaim the memory used for representing \prog{xs} as it is deconstructed by the \prog{match} expression.

\begin{figure}[H]
  \scriptsize
  \begin{tllisting}
program lappend {A : L} (xs : llist A) : llist A $⊸$ llist A =
ln ys $⇒$ match xs with
  | lnil $⇒$ ys
  | lcons x xs $⇒$ lcons x (lappend _ xs ys)
  end
  \end{tllisting}
  \caption{Program for Appending Linear Lists}
  \label{lappend}
\end{figure}

After a program has been defined, theorems regarding its properties can be proven at the logical level using the \prog{logical} keyword. \Cref{lappend-len} shows a proof~\footnote{The notation \progtiny{rew [x, p $⇒$ A] P in H} is the propositional equality eliminator $\idenElim{[x,p]A}{H}{P}$.} that the logical length of two lists appended together by \prog{lappend} is equal to the sum of their individual lengths. Notice that the \prog{llen} function here is a logical specification: it cannot be used at the program level for actual computations. If \prog{llen} were to exist relevantly at the program level, the element \prog{hd} dropped without usage in the \prog{lcons} case would cause memory leakage. Basically, terms declared with the \prog{logical} keyword are not subject to linear type checking and are pruned during the erasure phase of the compiler.

\begin{figure}[H]
  \scriptsize
  \begin{tllisting}
logical llen {A : L} (xs : llist A) : nat =
  match xs with
  | lnil $⇒$ 0
  | lcons hd tl $⇒$ 1 + llen _ tl
  end

logical lappend_llen (A : L) (xs ys : llist A) :
  llen _ (lappend _ xs ys) $≡$ (llen _ xs) + (llen _ ys) =
  match xs as xs0 in llen _ (lappend _ xs0 ys) $≡$ llen _ xs0 + llen _ ys with
  | lnil $⇒$ refl
  | lcons x xs0 $⇒$
    rew [ n, _ $⇒$ S (llen _ (lappend _ xs0 ys)) $≡$ S n ]
    lappend_llen _ xs0 ys in refl
  end
  \end{tllisting}
  \caption{Logical Proofs Relating Append and Length}
  \label{lappend-len}
\end{figure}

The example that was just shown is a form of \textit{extrinsic} verification. In this style of verification, an unverified program is first written using standard programming techniques. After the program has been fully constructed, its properties are then stated and proven as theorems external to the program.

Taking advantage of dependent types, programs can also be verified in an \textit{intrinsic} manner where data and proofs are tightly integrated. Consider length indexed linear vector defined in \Cref{lvec}. The constructors \prog{lNil} and \prog{lCons} of this inductive type carry irrelevant proofs that the indexing natural number \prog{n} accurately characterizes the length of the constructed vector. So if a vector is known to be of type \prog{lvec n A}, we can trust that its length must be \prog{n}.

\begin{figure}[H]
  \scriptsize
  \begin{tllisting}
inductive lvec (n : nat) (A : L) : L =
| lNil  of {e : 0 $≡$ n}
| lCons of {n0 : nat} {e : S n0 $≡$ n} (hd : A) (tl : lvec n0 A)
  \end{tllisting}
  \caption{Linear Length Indexed Vectors}
  \label{lvec}
\end{figure}

A program for appending linear vectors is given in \Cref{vappend}. We can see immediately from the type of the output \prog{lvec (m + n) A} that its length must be exactly the sum of the lengths of its inputs. Compared to the extrinsic approach, intrinsic verification can help to guide the process of program construction itself as the tightly integrated proofs serve as precise interfaces that rule out incorrect programs.

\begin{figure}[H]
  \scriptsize
  \begin{tllisting}
program vappend {m n : nat} {A : L} (xs : lvec m A) : lvec n A $⊸$ lvec (m + n) A =
ln ys $⇒$ match xs with
  | lNil e $⇒$ rew [ m0, _ $⇒$ lvec (m0 + n) A ] e in ys
  | lCons n0 e hd tl $⇒$
    rew [ m0, _ $⇒$ lvec (m0 + n) A ] e in
    lCons (n0 + n) refl hd (vappend _ _ _ tl ys)
  end
  \end{tllisting}
  \caption{Program for Appending Linear Vectors}
  \label{vappend}
\end{figure}

The computational relevancy mechanism of TLL allows irrelevant constructor arguments to be safely erased. For \prog{lvec} in particular, constructor arguments surrounded by braces are erased. The structure of \prog{lvec} after erasure is identical to \prog{llist}. If \prog{lvec} was defined verbatim in Coq, then the \prog{n0} argument of \prog{lCons} would not be erased because \prog{nat} is in the \prog{Set} universe. 

\subsection{Sort-Polymorphism}
In TLL, non-linear types and linear types are unambiguously grouped in sorts $\Un$ and $\Ln$ respectively as shown through the sort uniqueness theorem (\Cref{sort-unique}). This means that multiple versions of equivalent functions may need to be defined at different sorts, causing large amounts of code duplication. Consider the polymorphic identity functions shown in \Cref{mono-id}. The first function \prog{idU} is polymorhpic over non-linear types and the second \prog{idL} is polymorphic over linear types.

\begin{figure}[H]
  \scriptsize
  \begin{tllisting}
program idU {A : U} (x : A) : A = x
program idL {A : L} (x : A) : A = x
  \end{tllisting}
  \caption{Sort Monomorphic Identity Functions}
  \label{mono-id}
\end{figure}

In order to reduce code duplication, we implement sort-polymorphism in the TLL compiler. Toplevel declarations are allowed to quantify over sorts using sort variables. We refer to these sort quantified declarations as \textit{sort-polymorphic schemes}. \Cref{poly-id} shows how a sort-polymorphic identity function can be defined. The \prog{id<s>} scheme is parameterized by sort variable \prog{s} which is then used by \prog{Type<s>} to refer to the sort of type \prog{A} generically.

\begin{figure}[H]
  \scriptsize
  \begin{tllisting}
program id<s> {A : Type<s>} (x : A) : A = x
  \end{tllisting}
  \caption{Sort Polymorphic Identity Function}
  \label{poly-id}
\end{figure}

It is important to note that schemes are not proper terms in TLL. Instead, the compiler attempts to instantiate the sort parameters of schemes with all possible combinations of $\Un$ and $\Ln$. The instantiated instances that pass type checking are elaborated into sort-monomorhpic TLL terms. Conversely, instances that do not pass type checking are pruned. For schemes such as \prog{id} where both instantiated instances are well-typed, the compiler will essentially derive \prog{idU} and \prog{idL} automatically and apply the correct version depending on the sort of its argument.

Sort-polymorhpic schemes can also be used for deriving inductive types. \Cref{poly-list} defines a \prog{list<s,t>} inductive type scheme whose elements are of sort \prog{s} and that itself is of sort \prog{t}. The \prog{llist} type presented in \Cref{inductive} can be viewed as the instantiated instance \prog{list<L,L>} where the elements of the list are linear and the list itself is also linear. An interesting instance that could be obtained from instantiating the \prog{list<s,t>} scheme is \prog{list<L,U>}. The \prog{cons} constructor of \prog{list<L,U>} is unsound as it enables the duplication of linear resources by first packing them into non-linear lists. To prevent such unsound situations from occurring, the \prog{cons} constructor of \prog{list<L,U>} is pruned, leaving \prog{list<L,U>} with only \prog{nil} as its constructor. In general, constructors of non-linear inductive types may only take arguments which are also non-linear. The constructors of scheme instances that do not satisfy this criteria are pruned. 

\begin{figure}[H]
  \scriptsize
  \begin{tllisting}
inductive list<s,t> (A : Type<s>) : Type<t> =
| nil
| cons of (hd : A) (tl : list<s,t> A)
  \end{tllisting}
  \caption{Sort-Polymorhpic Lists}
  \label{poly-list}
\end{figure}

\Cref{poly-len} presents a length function for sort-polymorphic lists. Notice that unlike the logical \prog{llen} shown in \Cref{lappend-len}, the new \prog{len<s,t>} function also returns back the original input list paired with its length. This makes \prog{len<s,t>} sound for use as a computationally relevant program as the aforementioned memory leakage problem regarding \prog{hd} is no longer possible. Furthermore, we can prove at the logical level that the list returned by \prog{len<s,t>} is indeed equal to its input. Due to the fact that the \prog{len\_id<s,t>} theorem here is also sort-polymorphic, this single proof suffices to verify \prog{len<s,t>} for all valid sort variations of lists.

\begin{figure}[H]
  \scriptsize
  \begin{tllisting}
program len<s,t> {A : Type<s>} (xs : list<_,t> A) : nat $⊗$ list<_,t> A =
  match xs with
  | nil $⇒$ $⟨$0, nil$⟩$
  | cons hd tl $⇒$
    match len _ tl with
    | $⟨$n, tl$⟩$ $⇒$ $⟨$S n, cons hd tl$⟩$
    end
  end

logical len_id<s,t> {A : Type<s>} (ls : list<_,t> A) : ls $≡$ snd _ _ (len _ ls) =
  match ls as ls0 in ls0 $≡$ snd _ _ (len _ ls0) with
  | nil $⇒$ refl
  | cons x xs $⇒$ ...
  end
  \end{tllisting}
  \caption{Sort-Polymorhpic Length (Excerpt)}
  \label{poly-len}
\end{figure}

\section{Related Work}
\subsection{Computational Relevancy}
Over the years, a number of mechanisms have been proposed for specifying computational relevancy and program extraction. The Dependent ML (DML) of Xi~\cite{dml} uses a stratified language where the static fragment is irrelevant and the dynamic fragment is relevant. A special class of indexed singleton types carries information between the statics and the dynamics. Miquel~\cite{miquel01} introduces the Implicit Calculus of Constructions (ICC) which extends the standard Calculus of Constructions of Coquand~\cite{cc} with intersection types. Intersection types allow implicitly quantified terms to be instantiated with hypothetical arguments that are not explicitly present in the syntax tree. ICC programs essentially never carry around irrelevant terms in the first place. Due to the fact that these instantiating arguments must be synthesized spontaneously without additional information from the syntax tree, type checking for ICC is undecidable. Barras and Bernardo~\cite{barras08} develop a decidable variation of ICC (ICC*) by requiring explicit instantiations for implicit quantifiers. An erasure procedure is then carried out to remove the arguments of implicit instantiations.

From a computational relevancy perspective, TLL can be viewed as an integration of DML style stratification and ICC* style implicit quantification. On the one hand, stratification distinguishes proofs from programs in a straightforward manner that enforces types and assumed axioms to always be irrelevant. Additionally, the operational semantics of the two levels can be tailored to better facilitate reasoning and computation independently of each other. On the other hand, implicit quantifications allow for Martin-L\"{o}f style dependency~\cite{martinlof} which is more expressive than DML style dependency.

\subsection{Combining Dependency and Linearity}
Linear types are a class of type systems inspired by Girard's sub-structural Linear Logic~\cite{girard}. Girard notices that the weakening and contraction rules of classical logic when restricted carefully, give rise to a new logical foundation for reasoning about resources. Wadler~\cite{wadler1990} applies an analogous restriction to variable usage in simple type theory, leading to the development of linear type theory where expressions respect resources. Programming languages featuring linear types~\cite{linear-haskell} or affine-like types~\cite{rust} have been implemented, allowing programmers to write resource safe software in practical applications.

Work has been done to enrich linear type theories with dependent types. Cervesato and Pfenning~\cite{llf} extend LF~\cite{lf} with linear types as LLF, being first to demonstrate that dependency and linearity can coexist within a type theory. The ATS programming language~\cite{ats} extends DML style dependent types with linear types to facilitate safe effectful programming. V\'{a}k\'{a}r~\cite{vakar14} presents a dependent linear type theory ILDTT with an underlying categorical semantics. Krishnaswami et al.~\cite{neel15} introduce a dependent linear type theory LNL$_{D}$ based on Benton's~\cite{benton1994} prior work of mixed linear and non-linear calculus. Although LNL$_{D}$ also employs a stratified type system, the stratification here is not used for specifying computational relevancy. Instead, the stratification of LNL$_{D}$ is used for separating linear types from non-linear types and Miquel style intersection types are used for encoding computational relevancy.

The works described so far all prohibit types from depending on linear terms in order to prevent resource duplication within types. Luo and Zhang~\cite{luo} are the first to describe a system where linear dependency is allowed. To accomplish this, they introduce the notion of \textit{essential linearity} which points out that types are hypothetical entities so linear terms occurring inside types should not contribute to overall resource consumption. Our work is inspired by essential linearity and also supports linear dependency which allows one to prove theorems regarding linear programs.

Based on initial ideas of McBride~\cite{nothing}, Atkey's QTT~\cite{qtt} uses semi-ring annotations to track variable occurrence, simulating computational relevancy, linear types and affine types within a unified framework. The heap semantics analysis of Choudhury et al.~\cite{choudhury21} show that QTT requires a form of reference counting to garbage collect unneeded resources at runtime. The soundness theorems for our heap semantics guarantee TLL programs to be memory clean without runtime garbage collection.

\section{Conclusion and Future Work}
TLL is a two-level dependent type theory that aims to characterize the nature of proofs and programs faithfully. Hosting a structural type system that is reminiscent of Martin-L\"{o}f type theory~\cite{martinlof}, the logical level derives hypothetical objects such as proofs and types which are computationally irrelevant. The program level uses the types and proofs derived at the logical level to realize a Linear Logic~\cite{girard} inspired type system. Programs constructed using the program level rules can be freely reflected into the logical level for hypothetical reasoning. We develop an erasure procedure for removing irrelevant terms occurring inside programs and show that programs extracted this way maintain computational productivity. Through a heap semantics analysis we prove that extracted TLL programs run memory clean.

We plan to investigate what additional extensions are possible with the additional flexibility afforded to us by TLL's stratified design. Our ultimate goal is to strive towards a framework that synergistically unites theorem proving and practical programming.

\bibliography{reference}
\end{document}